\documentclass{ws-procs975x65}
\def\be{\begin{equation}}
\def\ee{\end{equation}}
\def\bq{\begin{eqnarray}}
\def\eq{\end{eqnarray}}
\def\beq{\begin{eqnarray*}}
\def\eeq{\end{eqnarray*}}

\begin{document}

\markboth{Antoniadis et al}{Braneworld Singularities}

\wstoc{Braneworld Singularities}{Antoniadis et al}

\title{Braneworld cosmological singularities}
\author{Ignatios Antoniadis$^{1,\ast}$\footnote{On leave from CPHT (UMR CNRS 7644) Ecole Polytechnique, 91128 Palaiseau Cedex, France},
Spiros Cotsakis$^{2,\dag}$, Ifigeneia Klaoudatou$^{3,\ddag}$}
\address{$^1$ Department of Physics, CERN -- Theory Division, CH-1211, Geneva 23, Switzerland,\\
$^{2,3}$ University of the Aegean, Karlovassi, 83 200 Samos, Greece\\
\email{ignatios.antoniadis@cern.ch$^\ast$,\,\, skot@aegean.gr$^\dag ,$\,\, iklaoud@aegean.gr $^\ddag$}}

\bodymatter

The purpose of this brief report is to present some results of our
on-going project on the asymptotic behaviour of braneworld-type
solutions on approach to their possible finite `time' singularities.
Cosmological singularities in such frameworks have served as  means
to attack the cosmological constant problem (see \cite{savas}
 and references therein). The main mathematical tool
of our analysis is the method of asymptotic splittings introduced in
Ref. \cite{cb}.

Below we study a model consisting of a $3-$brane configuration
embedded in a five dimensional bulk space with a scalar field being
minimally coupled to the bulk and conformally coupled to the fields
restricted on the brane. The total action is taken to be
$S_{total}=S_{bulk}+S_{brane}$, where $$S_{bulk}=\int d^{4}x
dY\sqrt{g_{5}}\left(\frac{R}{2\kappa^{2}_{5}}-
\frac{\beta}{2}(\nabla\phi)^{2}\right),\,\, S_{brane}=-\int
d^{4}x\sqrt{g_{4}}f(\phi),\,\, \textrm{at}\,\,Y=Y_{\ast},$$ with $Y$
denoting the fifth bulk dimension, $\kappa^{2}_{5}=M_{\ast}^{-3}$,
$M_{\ast}$ being the five dimensional Planck mass and $f(\phi)$ is
the tension of the brane depending on the scalar field $\phi$. We
assume a bulk metric of the form
$ds^{2}=a^{2}(Y)d\tilde{s}^{2}+dY^{2}$, where $d\tilde{s}^{2}$ is
the four dimensional flat, de Sitter or anti-de Sitter metric. Then
varying the above action we obtain the field equations: \be
\label{feq1}
\frac{a'^{2}}{a^{2}}=\frac{\beta\kappa^{2}_{5}\phi'^{2}}{12}+\frac{k
H^{2}}{a^{2}} \ee \be \label{feq2}
\frac{a''}{a}=-\frac{\beta\kappa^{2}_{5}\phi'^{2}}{4}, \quad
\phi''+4\frac{a'}{a}\phi'=0, \ee where $k=0,1$ or $-1$, and $H^{-1}$
is the de Sitter curvature radius. Assuming further that the
unknowns are invariant under a $Y\rightarrow -Y$ symmetry and
solving the field equations on the brane we may express the solution
in the form \be \label{bound1}
a'(Y_{\ast})=-\frac{\kappa_{5}^{2}}{6}f(\phi(Y_{\ast}))a(Y_{\ast}),
\quad \phi'(Y_{\ast})=\frac{f'(\phi(Y_{\ast}))}{2\beta}. \ee We now
apply the method of asymptotic splittings  to look for the possible
asymptotic behaviours of the general solution. Setting $x=a$,
$y=a'$, $z=\phi'$, where the differentiation is considered with
respect to $\Upsilon=Y-Y_{s}$ ($Y_{s}$ being the position of the
singularity), the field equations (\ref{feq2}), become the following
system of ordinary differential equations: \be \label{syst1}
x'=y,\,\, y'=-\beta Az^{2}x,\,\, z'=-4yz/x, \ee where
$A=\kappa^{2}_{5}/4$. Hence, we have the vector field
$\mathbf{f}=(y,-\beta Az^{2}x,-4yz/x)^{\intercal}$. Equation
(\ref{feq1}) does not include any terms containing derivatives with
respect to $\Upsilon$; it is the constraint equation of the above
system. In terms of the new variables, the constraint has the form
\be \label{constraint1} y^{2}/x^{2}=A\beta/3 z^{2}+k H^{2}/x^{2}.
\ee Substituting the forms $(x,y,z)=(\alpha\Upsilon^{p},\gamma
\Upsilon^{q},\delta\Upsilon^{r}),$ with $(p,q,r)\in\mathbb{Q}^{3}$
and $(\alpha,\gamma,\delta)\in \mathbb{C}^{3}-\{\mathbf{0}\}$, in
the dynamical system (\ref{syst1}), we seek to determine the
 possible \emph{dominant balances}  in the neighborhood of the singularity, that is pairs of the form
$\mathcal{B}=\{\mathbf{a},\mathbf{p}\}$, where
$\mathbf{a}=(\alpha,\gamma,\delta)$ and $\mathbf{p}=(p,q,r)$.
For our system we find:
\bq \label{sing}
\mathcal{B}_{1}&=&\{(\alpha,\alpha/4,\sqrt{3}/4\sqrt{A\beta}),(1/4,-3/4,-1)\}\\
\label{nonsing} \mathcal{B}_{2}&=&\{(\alpha,\alpha,0),(1,0,-1)\}\\
\mathcal{B}_{3}&=&\{(\alpha,0,0),(0,-1,-1)\}. \eq
Since (\ref{syst1}) is a weight-homogeneous system, the scale invariant
solutions given by the above balances are exact solutions of the system. The balance
$\mathcal{B}_{1}$ satisfies the constraint equation (\ref{constraint1})
only for $k=0$, corresponding thus to a general solution
for a flat brane, whereas
$\mathcal{B}_{2}$ corresponds to a particular solution for a
curved brane since it satisfies Eq. (\ref{constraint1}) for $k\neq 0$ and
$\alpha^{2}=k H^{2}$. Finally the balance $\mathcal{B}_{3}$ represents a
static universe conformal to Minkowski space and will not be analyzed further.

Next we calculate the Kowalevskaya exponents, i.e., the
eigenvalues of the matrix given by
$\mathcal{K}=D\mathbf{f}(\mathbf{a})-\textrm{diag}(\mathbf{p})$;
for $\mathcal{B}_{1}$ we find that
$\textrm{spec}(\mathcal{K})=\{-1,0,3/2\}$, whereas for
$\mathcal{B}_{2}$,  $\textrm{spec}(\mathcal{K})=\{-1,0,-3\}.$ These
exponents correspond to the indices of the series coefficients
where arbitrary constants first appear. The $-1$ exponent
signals the arbitrary position of the singularity, $Y_{s}$.
Since we have two non-negative integer eigenvalues the solution we
are constructing will be a general solution (full number of
arbitrary constants).

Let us now focus on each of the two possible balances separately
and build series expansions in the neighborhood of the
singularity. For the first balance, we substitute in the system
(\ref{syst1}) the  series expansions
$\mathbf{x}=\Upsilon^{\mathbf{p}}(\mathbf{a}+
\Sigma_{j=1}^{\infty}\mathbf{c}_{j}\Upsilon^{j/s}),$ where
$\mathbf{x}=(x,y,z)$, $\mathbf{c}_{j}=(c_{j1},c_{j2},c_{j3})$, $s$ is
the least common multiple of the denominators of positive
eigenvalues (here $s=2$), and we arrive at the asymptotic solution
\be
x=\alpha\Upsilon^{1/4}+\frac{4}{7}c_{32}\Upsilon^{7/4}+\cdots,
\quad
y=x',\quad z=\frac{\sqrt{3}}{4\sqrt{A}}\Upsilon^{-1}-
\frac{4\sqrt{3}}{7\alpha\sqrt{A\beta}}c_{32}\Upsilon^{1/2}+\cdots.
\ee
The last step is to check if, for each $j$ satisfying
$j/s=\rho$ with $\rho$ a positive eigenvalue corresponding to an
eigenvector $\mathbf{v}$ of the $\mathcal{K}$ matrix, the compatibility conditions hold, i.e.
$\mathbf{v}^{\top}\cdot \mathbf{P}_{j}=0$, where $\mathbf{P}_{j}$ are polynomials in
$\mathbf{c}_{i},\ldots, \mathbf{c}_{j-1}$ given by
$\mathcal{K}\mathbf{c}_{j}-(j/s)\mathbf{c}_{j}=\mathbf{P}_{j}.$ Here the
corresponding relation $j/2=3/2$ is valid only for $j=3$ and the
compatibility condition indeed holds. We therefore  conclude that
near the  singularity at finite distance $Y_{s}$ from the brane, the
asymptotic forms of the variables are $a\rightarrow 0$,
$a'\rightarrow\infty$, $\phi'\rightarrow \infty.$ This is exactly
the asymptotic behaviour of the solution found previously by
Arkani-Hammed \emph{et al} in Ref. \cite{savas}.

However, the previous behaviour is not the only possible one. The second balance has two distinct negative Kowalevskaya exponents and we
therefore expect to find
an infinite expansion of a \emph{particular} solution around the
presumed singularity at $Y_{s}$. Expanding the variables in series
with descending powers of $\Upsilon$, in order to meet the two
arbitrary constants occurring $j=-1$ and $j=-3$, and substituting back
in the system (\ref{syst1}) we find the forms
\be
x=\alpha\Upsilon+c_{-1\,1}+\cdots,\quad
y=\alpha+\cdots,\quad
z=c_{-3\,3}\Upsilon^{-4}+\cdots
\ee
Therefore as $\Upsilon\rightarrow 0$, or equivalently as $S=1/\Upsilon\rightarrow \infty$,
we have that $a\rightarrow \infty$, $a'\rightarrow
\infty$ and $\phi'\rightarrow \infty$.

We thus conclude that there exist two possible outcomes for these
braneworld models, the dynamical behaviours of which strongly depend
on the spatial geometry of the brane. For a flat brane the model
experiences a finite distance singularity through which all the
vacuum energy decays, whereas for a de Sitter or anti-de Sitter
brane the singularity is now located at an infinite distance. We can
choose the coupling such that to allow only for that behaviour met
in flat solutions and, in fact, the particular form of the coupling
used by Arkani-Hammed \emph{et al} in \cite{savas} is the only
choice to make this possible. This easily follows by using equations
(\ref{bound1}) and solving the Friedmann equation
(\ref{feq1}) on the brane for $kH^{2}$, i.e.
$$kH^{2}=\frac{a^{2}(Y_{\ast})\kappa^{2}_{5}}{4}
\left(\frac{\kappa_{5}^{2}}{9}f^{2}(\phi)-
\frac{f'^{2}(\phi)}{4\beta^{2}}\right).$$ Clearly then $k$ is
identically zero if and only if
$f'(\phi)/f(\phi)=(2\beta/3)\kappa_{5}$, or equivalently, if and
only if  $f(\phi)\propto e^{(2\beta/3)\kappa_{5}\phi}$ (Arkani-Hammed
\emph{et al} in \cite{savas} have $\beta=3$). By working with other
couplings we can allow for non-flat, maximally symmetric solutions
to exist and avoid in this way having the singularity at a finite
distance away from the position of the brane.

\vspace{0.5cm} I.A. was supported in part by the European Commission
under the RTN contract MRTN-CT-2004-503369, while S.C. and I.K. were
supported by the joint E.U. and Greek Ministry of Education grants
`Pythagoras' and `Herakleitos' respectively. S.C. and I.K. are very
grateful to CERN-Theory Division, where part of their work was done,
for making their visits there possible and for allowing them to use
its excellent facilities. This work of I.K. represents a partial
fulfilment of the PhD requirements, University of the Aegean.

\vfill

\end{document}